\documentclass{aastex61}
\usepackage{amsmath}
\usepackage{longtable}
\received{today}
\submitjournal{ApJ}

\shorttitle{Tracers of protoplanetary disk gas mass}
\shortauthors{Molyarova et al.}

\begin{document}

\title{Gas mass tracers in protoplanetary disks: CO is still the best}

\correspondingauthor{Tamara Molyarova}
\email{molyarova@inasan.ru}

\author{Tamara Molyarova}
\affil{Institute of Astronomy, Russian Academy of Sciences, 48 Pyatnitskaya St., Moscow, 119017, Russia}
\affiliation{Moscow Institute of Physics and Technology (State University), 9 Institutskiy per., Dolgoprudny, Moscow Region, 141700, Russia}

\author{Vitaly~Akimkin}
\affiliation{Institute of Astronomy, Russian Academy of Sciences, 48 Pyatnitskaya St., Moscow, 119017, Russia}

\author{Dmitry~Semenov}
\affiliation{Max Planck Institute for Astronomy, K{\"o}nigstuhl 17, 69117 Heidelberg, Germany}

\author{Thomas~Henning}
\affiliation{Max Planck Institute for Astronomy, K{\"o}nigstuhl 17, 69117 Heidelberg, Germany}

\author{Anton~Vasyunin}
\affiliation{Max Planck Institute for Extraterrestrial Physics, Giessenbachstrasse, 85748 Garching, Germany}
\affiliation{Ural Federal University, Ekaterinburg, 620083, Russia}

\author{Dmitri~Wiebe}
\affiliation{Institute of Astronomy, Russian Academy of Sciences, 48 Pyatnitskaya St., Moscow, 119017, Russia}

\begin{abstract}
Protoplanetary disk mass is a key parameter controlling the process of planetary system formation. CO molecular emission is often used as a tracer of gas 
mass in the disk. In this study we consider the ability of CO to trace the gas mass over a wide range of disk structural parameters and search for 
chemical species that could possibly be used as alternative mass tracers to CO. Specifically, we apply detailed astrochemical modeling to a large set 
of models of protoplanetary disks around low-mass stars, to select molecules with abundances correlated with the disk mass and being relatively 
insensitive to other disk properties. We do not consider sophisticated dust evolution models, restricting ourselves with the 
standard astrochemical assumption of $0.1~\mu $m dust. We find that CO is indeed the best molecular tracer for total gas mass, despite the fact that it 
is not the main carbon carrier, provided reasonable assumptions about CO abundance in the disk are used. Typically, chemical reprocessing lowers the abundance of CO by a factor of 3, compared to the case of photo-dissociation and freeze-out as the only ways of CO depletion. On average only 13\% C-atoms reside in gas-phase~CO, albeit with variations from 2 to 30\%. CO$_2$, H$_2$O and H$_2$CO can potentially serve as alternative mass tracers, the latter two being only applicable if disk structural 
parameters are known.
\end{abstract}

\keywords{protoplanetary disks --- astrochemistry --- circumstellar matter}

\section{Introduction}

A young Sun-like star at its early evolutionary stage is generally surrounded by a flattened rotating gas and dust structure, which is commonly referred to 
as a 
protoplanetary disk even though evidence of planet formation in these disks is still mostly circumstantial 
\citep[see, e.g., reviews by][]{2011ARA&A..49...67W,2015arXiv150906382A}. While the exact scenario of planet formation remains debated, the mass of the 
protoplanetary disk is thought to play a key role in this process \citep{2012A&A...541A..97M, 2015A&A...582A.112B}. Therefore, estimating accurate disk 
masses from observations is one of the most important but also most challenging tasks in studies of formation and early evolution of planetary systems.

There are two different approaches to measuring disk masses. The most common way is to observe the dust continuum emission. Provided this 
emission is optically thin (which is expected at sub-millimeter and millimeter wavelengths), one can calculate the total dust mass in the disk 
\citep{2005ApJ...631.1134A,2011ARA&A..49...67W} from the integrated continuum flux. Then, to convert the dust mass into a total disk mass, standard 
interstellar 
gas-to-dust ratio of 100 can be adopted \citep{1978ApJ...224..132B}. However, to apply this method one needs to know the dust temperature in the disk, 
because the same flux at a given wavelength may come both from larger amount of cold dust or from a smaller amount of warm dust. Such a mass derivation 
requires assumptions about dust optical properties, which are not well constrained for dust grains in protoplanetary disks and also may change as dust 
grows at early stages of planet formation \citep{2006ApJ...636.1114D,2011ppcd.book..114H}. Furthermore, it is far from obvious that interstellar 
gas-to-dust mass ratio is valid for protoplanetary disks. \citet{2014MNRAS.444..887D} and \citet{2017ApJ...838..151T} suggest that masses of young 
disks derived from dust observations may be underestimated.

The more direct way to measure gas mass in disks is to observe molecular emission lines. While disks mostly consist of molecular hydrogen and helium, 
these components either cannot be observed directly, or trace only a limited disk region \citep[see e.g.][]{2011A&A...533A..39C}, and other mass tracers are 
commonly employed, which require knowledge of a conversion factor from tracer mass to H$_2$ mass. In this respect, hydrogen deuteride (HD) seems to be a 
good candidate \citep{2013Natur.493..644B,2016ApJ...831..167M}, as it is well coupled to molecular hydrogen and also has a relatively high abundance of $3 
\cdot 10^{-5}$ relative to H$_2$, although using it as a mass tracer may lead to a factor of ~3 uncertainty \citep{2017arXiv170507671T}. Currently, the 
only facility to produce new HD measurements will be SOFIA with its forthcoming HIRMES spectrograph.

A widely used molecular mass tracer in disks is carbon monoxide (CO), as it is one of the most abundant species in the interstellar and circumstellar material and possesses rotational transitions that are easy to observe. The hydrogen mass determination using CO lines is usually applied to diffuse clouds \citep{1997ApJ...491..615G,2013ARA&A..51..207B}, but now it is also commonly used for protoplanetary disks \citep{WB2014,2016ApJ...828...46A, WMP2016,2016A&A...594A..85M}. But the CO-based disk mass determination method has some recognized problems, too. First, CO emission is optically thick, which can be circumvented by observing CO isotopologues $^{13}$CO, C$^{18}$O, and C$^{17}$O, and also $^{13}$C$^{18}$O, which has been recently observed in the TW~Hya disk \citep{2017NatAs...1E.130Z}. Their lines are typically optically thin as they are less abundant \citep{2011ApJ...740...84Q}. The second problem is that CO along with its isotopologues freezes out in cold dense regions of the disk midplane and gets non-observable \citep{2013ChRv..113.9016H}.

Given the lack of good alternatives, CO observations are still vital for the determination of protoplanetary disk mass. However, its application for that 
purpose should be validated with theoretical models comprising both radiation transfer and chemistry. \cite{WB2014} constructed a model of CO distribution 
over a selection of parameterized protoplanetary disks, assuming that CO is frozen in those regions of the disk where the temperature is below 20~K and is 
photo-dissociated above the vertical H$_2$ column density of $1.3\times10^{21}$ cm$^{-2}$. In the remaining part of the disk the CO abundance relative to 
H$_2$ was assumed to be equal to the so-called `interstellar' value, $10^{-4}$ \citep{france}. \cite{WB2014} pointed out that under these assumptions, CO 
mass in this warm molecular zone represents a significant fraction of the total CO mass and thus traces the bulk of the disk gas mass. However, the 
CO-based disk gas masses derived so far seem to be systematically smaller than masses derived from dust observations. ALMA disk surveys in Lupus, 
Chamaeleon, $\sigma$ Orionis and other star-forming regions by \cite{2016ApJ...828...46A,2017AJ....153..240A} based on the same theoretical model suggest 
very high dust-to-gas mass ratios up to $1 / 10$.

\cite{WB2014} suggested that the difference between the CO-based disk mass and the dust-based disk mass can be both real (reflecting low gas-to-dust mass 
ratios) and related to an uncertainty in the CO-to-H$_2$ abundance ratio, if, for example, some other CO depletion pathways are more efficient. This 
possibility was explored by \cite{2016A&A...594A..85M} with the isotope-selective chemical model by means of artificially reduced carbon abundance. 
However, they only considered simple hydrogenation surface processes, which may somewhat decrease gas-phase CO abundance. They noted that the uncertainty 
in the CO-based gas mass estimate can be reduced if some information on the disk structure is available. A continuation of this study was presented by 
\cite{Miotello2017}, who confirmed that simplified surface chemistry may be the cause of discrepancy of disk masses estimated using CO and dust observations.

A more detailed surface chemistry model was utilized in the work of \cite{Yu2016}. These authors considered CO evolution in the model of an 
accretion-heated disk and found that even inside a CO snowline a gas-phase CO abundance can be reduced significantly due to conversion of CO into other 
less volatile species, like CO$_2$ and complex organic molecules (COMs), which is a common effect at various stages of star formation 
\citep{2012ApJ...751..105V}. However, only a limited number of disk models was considered in this study, and only an inner part of the disk (within 70 
au) was analyzed. Also, \cite{Yu2016} assumed that a significant dust evolution has already taken place in the disks, so that most dust grains have 
coagulated into larger particles, limiting efficiency of both CO freeze-out and surface conversion into other molecules. \cite{Yu2017} extended this study 
and came to the conclusion that the chemical depletion of CO may lead to significant underestimation of the disk gas masses.
 
In this study we intend to combine a large grid of disk physical models with a comprehensive chemical model, which includes detailed treatment of surface 
chemistry. We check the {\it degree} of reliability of CO as a gas mass tracer with realistic variations of disk parameters and the uncertainties related 
to CO chemical depletion. We perform an analysis of the carbon partition into the gas and solid species. The computed set of models is used to look for other 
possible tracers of disk gas mass. 

\section{Model}

The basis for our study is a set of $\sim 1000$ disk models (described in detail in Section~\ref{sec:models}), which allows studying the distribution 
of gas-phase and surface molecular abundances for a selection of disk structural parameters. We utilize a modified version of the ANDES code 
\citep{2013ApJ...766....8A} combined with the updated ALCHEMIC network \citep{2011ApJS..196...25S} to follow the evolution of molecular abundances.

\subsection{Physical Structure}

For the parameterization of the disk surface density we adopt the commonly used tapered power-law profile \citep{1998ApJ...495..385H}
\begin{equation}
     \Sigma\left(R\right) = \Sigma_0 \left(\frac{R}{R_{\rm c}}\right)^{-\gamma} e^{-\left(R/R_{\rm c}\right)^{2-\gamma}},
      \label{eq:sig}
\end{equation}
where $\Sigma_0$, g cm$^{-2}$, is the surface density normalization, $\gamma$ parameterizes the radial dependence of the disk viscosity, $R_{\rm c}$ is a 
characteristic radius of the disk. $\Sigma_0$ can be calculated from the total disk mass as
\begin{equation}
\Sigma_0 = \frac{M_{\rm disk}\left(2 - \gamma \right)}{2 \pi R_{\rm c}^{2}}.
\label{eq:sig0}
\end{equation}
So, the surface density radial profile depends on three parameters: $M_{\rm disk}$, $\gamma$, $R_{\rm c}$.

We assume that the disk is axially-symmetric and its vertical structure at each radius is determined by the hydrostatic equilibrium
\begin{eqnarray}
\frac{\partial P\left(R,z\right)}{\partial z} &=& -\rho \left(R,z \right) \frac{z G M_{\star}}{\left(R^2 +z^2\right)^{3/2}},\\
P &=& \frac{k_{\rm B} T\left(R,z\right)}{\mu m_{\rm p}} \rho \left(R,z \right),
\label{eq:vertdens}
\end{eqnarray}
where the disk temperature distribution $T\left(R,z\right)$ is calculated using a simple parameterization \citep{WB2014,isella2016ringed}
\begin{equation}
T\left(R,z\right) =
\begin{cases}
T_{\rm a}(R,z=z_{\rm q}) + \left[T_{\rm m}(R) - T_{\rm a}(R,z=z_{\rm q})\right]\left(\cos\frac{\pi z}{2 z_{\rm q}(R)}\right)^{2}, & \text{if 
$\left|z\right| < z_{\rm q}(R)$},\\
T_{\rm a}(R,z), & \text{if $\left|z\right| \geqslant z_{\rm q}(R)$}.
\label{eq:Tapprox}
\end{cases}
\end{equation}
Rather than adding disk atmosphere and midplane temperatures $T_{\rm m}$ and $T_{\rm a}$ to our set of parameters, we calculate them on the base of 
central luminosity source (star plus accretion region). The optically thin disk atmosphere can be easily ray-traced as dust thermal radiation does not 
contribute significantly into the source function. The dust temperature is determined from the thermal balance equation:
\begin{equation}
 \int\limits_0^\infty \kappa_\nu J_\nu d\nu =  \int\limits_0^\infty\kappa_\nu B_\nu (T_{\rm a}) d\nu
\end{equation}
where $\kappa_\nu$, cm$^{-1}$ is the dust absorption coefficient and $J_\nu$ is the mean intensity of radiation coming directly from the central star, 
accretion region and interstellar medium,
\begin{equation}
 J_\nu=J_{\nu,\rm s}+J_{\nu,\rm acc}+J_{\nu,\rm ISRF}=W_{\star}B_\nu(T_\star)e^{-\tau_{\nu,1}}+W_{\rm acc}B_\nu(T_{\rm acc})e^{-\tau_{\nu,1}}+\frac{1}{3}I_{\rm 
ISRF}(e^{-\tau_{\nu,2}}+e^{-\tau_{\nu,3}}+e^{-\tau_{\nu,4}}).
 \label{eq:radint}
\end{equation}
Here $W_\star=\frac{1}{4}R_\star^2/(R^2+z^2)$ and $W_{\rm acc}=\frac{1}{4}R_{\rm acc}^2/(R^2+z^2)$ represent the dilution factors for the radiation from 
the central star and accretion region, $R_\star$ and $T_\star$ are stellar radius and effective temperature. Accretion luminosity is simulated by the 
additional central source of irradiation with effective radius $R_{\rm acc}$ and effective temperature $T_{\rm acc}$, so that $L_{\rm acc}=4\pi R_{\rm 
acc}^2\sigma_{\rm SB}T_{\rm acc}^4=(3/2)\,GM_\star \dot{M}/R_\star$. We assume a typical accretion rate of $\dot{M}=10^{-8} M_{\odot}$~yr$^{-1}$ and 
$T_{\rm acc}=15\,000$\,K to calculate the effective radius of an accretion region, $R_{\rm acc}$. The interstellar radiation field is adopted from 
\citet{1983A&A...128..212M}.

The optical depth in \eqref{eq:radint}, $\tau_{\nu,k}=\int (\kappa_\nu + \sigma_\nu) dl$,  is calculated along four directions: to and from the star 
($k=1,2$), and up and down ($k=3,4$). Dust optical properties are computed from the Mie theory for astrosilicate grains \citep{1993ApJ...402..441L} 
assuming dust-to-gas ratio of 0.01. We use a spatial grid $R_n$ and $z_{nm}$ such that $z_{nm}/R_n$ is constant for a given $m$. This greatly reduces the 
computational cost for the ray-tracing procedure as all spatial points with fixed $m$-index are located on the same line.

We define $z_{\rm q}(R)$ as the height at which $T_{\rm a}$ has a maximum at a given $R$. Above $z_{\rm q}(R)$ dust temperature decreases with $z$ due to 
increasing distance from the star, the drop of temperature below $z_{\rm q}(R)$ is explained by the absorption. This also means that the source function 
starts to be dominated by the dust thermal infrared radiation and not by the radiation from star. In this regime the approximation for radiation intensity 
\eqref{eq:radint} is no longer valid. The temperature profile below $z_{\rm q}(R)$ is set according to \eqref{eq:Tapprox} with the midplane temperature 
adopted from the optically thick case of the two-layered model \citep{1997ApJ...490..368C,2001ApJ...560..957D}
\begin{equation}
T_{\rm m}^4(R) = \frac{1}{2} \varphi\left[ T_{\star}^4  \left(\frac{R_\star}{R}\right)^2 + T_{\rm acc}^4  \left(\frac{R_{\rm acc}}{R}\right)^2\right],
\label{eq:Tmprtr}
\end{equation}
 where $\varphi = 0.05$ is a flaring angle~\citep{2004A&A...421.1075D}, the factor $\frac{1}{2}$ reflect the fact that only half of the infrared radiation 
from the disk super-heated upper layer heats the midplane, and the other half is emitted outward. This midplane temperature differs from the disk effective 
temperature by a factor of $\sqrt[4]{2}$. Gas and dust temperatures are assumed to be equal. We use iterations to obtain self-consistent density and 
temperature structures, which is important for the ray-tracing procedure and calculation of the atmosphere temperature. However, in the calculation of the 
midplane temperature we keep the flaring angle fixed to avoid disk self-shadowing as this effect must be studied by means of a more sophisticated model. 
The adopted model for the disk physical structure enables us to perform astrochemical simulations for a large set of models while keeping the adequate 
level of accuracy. 

Values of the stellar radius $R_{\star}$, effective temperature $T_{\star}$ and stellar mass $M_{\star}$, used in the above parameterizations, are not 
independent. We take $M_{\star}$ as a primary parameter and calculate $R_{\star}$ and $T_{\star}$ from evolutionary tracks by \citet{2015A&A...577A..42B}, 
assuming a stellar age of 3~Myr. As their model is restricted to low-mass stars, we consider only T~Tauri-type stars with masses between 0.5 and 
1.4~$M_{\sun}$. Stellar and accretion luminosities are assumed to be constant during the chemical run. However, luminosity bursts may have a lasting 
impact on the disk chemical structure and are a topic of current research \citep{2015A&A...582A..41H, 2017arXiv170503946R}.  

Overall, we have four parameters, $M_{\rm disk}$, $\gamma$, $R_c$, and $M_{\star}$, which determine the physical structure of the disk. The spatial grid 
has 50~points in radial direction between 1~and 1000~au and 80~points in vertical direction from the midplane to $z/R=0.5$.

\subsection{Chemical Model}

The chemical structure is calculated with the modified version of the thermochemical non-equilibrium ANDES code \citep{2013ApJ...766....8A}. The system of 
chemical kinetics equations was updated to account for the gas-grain chemistry of the ALCHEMIC model \citep{2011ApJS..196...25S}. The gas-phase part was 
benchmarked with the NAHOON code \citep{2015ApJS..217...20W}, which resulted in fixing several non-critical bugs. Some rate coefficients were updated in 
accordance with the KIDA14 database. Overall, the chemical network contains 650 species involved in 7807 reactions. No isotope-selective chemistry is 
considered.

The original version of the ANDES code contains a module to calculate thermal balance of gas and dust. We replaced this module in the current study by a 
phenomenological setup for the thermal structure in order to reduce computational costs for modeling a large set of disks. However, the 2D radiation 
transfer (RT) of UV radiation was implemented in order to calculate photoreaction rates. The RT is based on the short characteristics method and allows 
for 
the radiation field calculation up to optical depths $\tau_{\rm UV}\sim10$. The MRN grain size distribution \citep{1977ApJ...217..425M} and dust-to-gas 
ratio of 0.01 are taken as the dust model for the RT modeling in the disk atmosphere. We calculate radiation transfer for the stellar and interstellar 
radiation field and also consider X-rays, cosmic rays (CR), and radioactive nuclides as sources of ionization. The spectrum of a central star is assumed 
to 
be a black body with an effective temperature corresponding to a given stellar mass plus an UV excess produced by accretion (as stated above). While the 
real stellar spectra may not be well represented by black body, in the context of our study only the relation between the visible and far-UV spectrum parts 
is important. The X-ray ionization rate is calculated on the base of \citet{2009ApJ...701..737B} model (their eq. [21]), assuming X-ray luminosity $L_{\rm 
X}=10^{30}$~erg~s$^{-1}$ and $T_{\rm X}=3$~keV. We use $\zeta_0=1.3\times10^{-17}$~s$^{-1}$ for unattenuated CR ionization rate and attenuation length of 
96~g~cm$^{-2}$ \citep{2000ApJ...543..486S} and $\zeta_{\rm RA}=6.5\times10^{-19}$~s$^{-1}$ for ionization rate by radioactive elements. 

The initial abundances are presented in Table~\ref{tab:iniab} and correspond to the `low-metals' case of \cite{1998A&A...334.1047L}, unless otherwise 
noted.
\begin{table}
  \centering  
  \caption{Initial abundances (relative to H-nuclei).}
  \label{tab:iniab}
  \begin{tabular}{l|cccccccc}
     Species & H$_2$ &     H &    He &   C$^+$   &    N      &    O      &    S$^+$  \\
         \hline
    Abundance & 0.499 & 0.002 &  0.09 &   $7.3\cdot10^{-5}$     &  $2.14\cdot10^{-5}$     &  $1.76\cdot10^{-4}$     &  $8\cdot10^{-8}$     \\
            \hline
            \hline
      Species &         Si$^+$ &    Fe$^+$ &    Na$^+$ &    Mg$^+$ &    P$^+$  &    Cl$^+$ &\\
      \hline
       Abundance &       $8\cdot10^{-9}$ &  $3\cdot10^{-9}$     &  $2\cdot10^{-9}$     &  $7\cdot10^{-9}$     &  $2\cdot10^{-10}$     &  $1\cdot10^{-9}$   
 
  &\\
  \end{tabular}
\end{table}

Apart from gas-phase processes, our chemical network includes surface reactions along with accretion and desorption. As many species tend to freeze out 
onto dust particles, which makes them unobservable, surface chemistry is crucial for such models. We adopt the ratio of diffusion energy to 
desorption energy of 0.5 and account for tunneling through reaction barriers. Single-layer surface chemistry is considered. Our treatment of surface 
chemistry includes reactive desorption with the efficiency of $1\%$, which means that in two-body surface reactions 99$\%$ of product species stay on the 
surface and 1$\%$ are released into gas, although more detailed treatment of reactive desorption is already available in some astrochemical models 
\citep{2017ApJ...842...33V}. The representative size of dust grains in grain chemistry treatment is assumed to be $0.1 \mu$m. In protoplanetary disks dust 
is supposed to grow, which reduces the available dust surface and slows down the surface chemistry. Also grown dust makes disk more transparent to stellar 
radiation. However, the theoretical description of dust evolution is a complicated task, especially taking into account such poorly studied effects as 
grain charging~\citep{2009ApJ...698.1122O,2015ARep...59..747A,2016ApJ...833...92I}. Therefore we neglect dust evolution effects and focus on varying disk 
macrophysical parameters.

\subsection{Set of Models}
\label{sec:models}

Like \citet{WB2014}, we create a grid of models with variable parameters responsible for the disk physical structure. As we have already mentioned, these 
parameters are disk mass $M_{\rm disk}$, stellar mass $M_{\star}$, effective radius of the disk $R_{\rm c}$, and power-law index $\gamma$. We vary them 
within bounds typical for protoplanetary disks around T~Tauri stars. Specific ranges for each parameter are presented in Table~\ref{tab:parameters}.

\begin{table}
  \centering  
  \caption{Range of parameters used in the set of models.}
  \label{tab:parameters}
  \begin{tabular}{l|l}
    Parameter & Value \\    
    \hline
    $M_{\rm disk}$ & $10^{-4}\div 10^{-1} M_{\odot}$ \\ 
    $M_{\star}$ & $0.5 \div  1.4M_{\odot}$ \\ 
    $R_{\rm c}$ & $30 \div  200$ au \\ 
    $\gamma$ & $0.5\div  1.5$ \\
  \end{tabular}
\end{table}

We calculate 2D distributions of 650 species for time moments of 0.5, 1, and 3~Myr and use these distributions to compute total species masses in the 
disk. 
In Section~\ref{sec:masstracer} we employ a random set of models, with every parameter being uniformly distributed inside the ranges (for disk mass it is 
$\log_{10}M_{\rm disk}$ that is distributed uniformly). In~Section~\ref{sec:gridofmodels} we use a grid of models with fixed nodes and present various cuts 
through the parameter space.

\section{Results}

\subsection{Potential gas mass tracers}
\label{sec:masstracer}

The list of molecules observed in protoplanetary disks is currently not very extensive, including CN, HCN, HNC, CS, SO, H$_2$CO, CCH, HC$_3$N, CH$_3$CN, 
C$_3$H$_2$, C$_2$H$_2$, OH, HCO$^+$, N$_2$H$^+$, CH$^+$, C$^+$, O, NH$_3$, CH$_3$OH \citep{1997A&A...317L..55D, 2011A&A...530L...2T, 2011A&A...535A.104D, 
2012A&A...537A..60C, 2013ApJ...765...34Q, 2015Natur.520..198O, 2016ApJ...823L..10W, 2016A&A...592A.124G, 2016A&A...591A.122S}. It would be interesting to find out if some of them could possibly serve as a gas mass tracer alternative to commonly used CO. Also, pending future discoveries of molecules in protoplanetary disks, we do not exclude from consideration other chemical species present in the model.

The total disk abundance of a prospective disk mass tracer $X_i = N_i/N_{\rm <H>}$ should be correlated with the disk mass, while being less dependent on other physical properties of the disk, such as the density profile, disk radius or properties of the central star. The correlation should not necessarily be linear, implying independence of the total abundance on the disk mass. In principle, the species abundance can be any monotonic function of the disk mass. Also, it should not vary significantly with time.

To assess the species potential applicability as a disk mass tracer, we need to introduce some quantitative gauges. For every $j$th disk model in the 
ensemble we calculate the total logarithmic abundance of $i$th species $x_{ij}=\log_{10} \left(X_{ij}\right)$. The value $\overline{x_i}$, averaged over the disk model ensemble, represents a typical (logarithm of) total disk abundance of an $i$th species. We employ its dispersion 
$s^2_i=\overline{(x_{ij}-\overline{x_i})^2}$ as a simple quantitative measure of suitability of $i$th species as a mass tracer. Further we refer to the
standard deviation $s$ as a `scatter parameter' to stress the limitation of the approach. First, the distribution of protoplanetary disks 
parameters is not well known and quite probably differs from the uniform distribution assumed during the disk ensemble generation. Second, the scatter may 
not reflect the non-linear scaling between disk and species mass. However, generally speaking, the smaller is the scatter, the better the species performs 
as a mass tracer.

Table~\ref{tab:variance} summarizes this simple statistical analysis and contains total disk abundances and scatter parameters for the model age 
of 3~Myr, sorted by increasing $s_i$. For other chemical ages the values of scatter parameter are about the same, slightly growing with time. We 
include components with average total disk abundance greater than $10^{-11}$, excluding ices and common species like H$_{2}$ and He. All the 
detected species listed above are present in the table as well. The species that are potentially observable with JWST but not with ALMA are marked with a 
star~($^{\star}$). Total disk abundances for ice species are given in the appendix in the table, analogous to Table~\ref{tab:variance}.

\begin{deluxetable*}{l|ll|l|ll|l|ll}
\tablecaption{Average total disk abundance $\overline{X_i}=10^{\overline{x_i}}$ and scatter $s_i$ for selected species\label{tab:variance}.}
\tablewidth{0pt}
\tablehead{
\colhead{Species} & \colhead{$\overline{X_i}$} & \colhead{$s_{i}$, dex} & 
\colhead{Species} & \colhead{$\overline{X_i}$} & \colhead{$s_{i}$, dex} & 
\colhead{Species} & \colhead{$\overline{X_i}$} & \colhead{$s_{i}$, dex} }
\startdata
N$_2$        & $  2.0 \times 10^{ -6}$  &  0.25 & C$_6$        & $  1.5 \times 10^{-11}$  &  0.45 & C            & $  4.9 \times 10^{ -6}$  &  0.64 \\
NH$_3$       & $  3.2 \times 10^{ -9}$  &  0.27 & HCN          & $  2.4 \times 10^{ -9}$  &  0.45 & O$_3$        & $  5.5 \times 10^{ -8}$  &  0.64 \\
CH$_3$       & $  3.6 \times 10^{-10}$  &  0.28 & Si           & $  1.7 \times 10^{-10}$  &  0.46 & O            & $  2.1 \times 10^{ -5}$  &  0.66 \\
CO           & $  8.0 \times 10^{ -6}$  &  0.28 & C$_4$H       & $ 10.0 \times 10^{-11}$  &  0.46 & C$_2$H$_6$   & $  2.3 \times 10^{ -9}$  &  0.66 \\
H$_3^+$      & $  2.2 \times 10^{-10}$  &  0.28 & $^{\star}$C$_2$H$_2$ 
                                                               & $  4.3 \times 10^{-10}$  &  0.47 & CH$_3^+$     & $  3.8 \times 10^{-11}$  &  0.73 \\
$^{\star}$CO$_2$ 
             & $  5.2 \times 10^{ -7}$  &  0.28 & Na           & $  1.9 \times 10^{-10}$  &  0.48 & C$_2$H       & $  6.6 \times 10^{-10}$  &  0.74 \\
H$_2$CO      & $  1.3 \times 10^{-10}$  &  0.29 & N$_2$H$^+$   & $  1.3 \times 10^{-11}$  &  0.48 & Na$^+$       & $  1.8 \times 10^{-10}$  &  0.74 \\
H$_3$O$^+$   & $  2.6 \times 10^{-10}$  &  0.30 & CS           & $  4.7 \times 10^{-10}$  &  0.49 & N            & $  1.5 \times 10^{ -6}$  &  0.75 \\
H$_2$CN$^+$  & $  1.6 \times 10^{-11}$  &  0.31 & Mg           & $  4.0 \times 10^{-11}$  &  0.49 & H$_5$C$_3$N  & $  2.2 \times 10^{-11}$  &  0.77 \\
H$_2$CS      & $  1.6 \times 10^{-11}$  &  0.32 & H$_2$C$_3$O  & $  4.7 \times 10^{-10}$  &  0.50 & CH$_2$       & $  5.6 \times 10^{-10}$  &  0.77 \\
NH$_2$       & $  3.5 \times 10^{-10}$  &  0.32 & MgH$_2$      & $  5.2 \times 10^{-10}$  &  0.50 & HC$_2$O      & $  6.9 \times 10^{-11}$  &  0.80 \\
SO$_2$       & $  2.1 \times 10^{-10}$  &  0.32 & HS           & $  2.1 \times 10^{-10}$  &  0.51 & N$_2$O       & $  1.2 \times 10^{-10}$  &  0.81 \\
HNO          & $  1.4 \times 10^{-10}$  &  0.33 & CH$_2$CO     & $  1.8 \times 10^{-10}$  &  0.51 & S$^+$        & $  4.9 \times 10^{ -9}$  &  0.81 \\
C$_3$        & $  1.3 \times 10^{ -8}$  &  0.34 & CH$_5$N      & $  2.5 \times 10^{-11}$  &  0.51 & HCCN         & $  5.0 \times 10^{-11}$  &  0.81 \\
C$_5$H$_3$   & $  2.5 \times 10^{-11}$  &  0.35 & $^{\star}$H$_2$O  
                                                               & $  4.0 \times 10^{ -7}$  &  0.51 & P$^+$        & $  1.1 \times 10^{-11}$  &  0.82 \\
SO           & $  2.7 \times 10^{ -9}$  &  0.37 & Cl           & $  1.8 \times 10^{-10}$  &  0.51 & Si$^+$       & $  5.1 \times 10^{-10}$  &  0.83 \\
C$_3$H       & $  1.1 \times 10^{-10}$  &  0.39 & CH$_3$CN     & $  2.0 \times 10^{-11}$  &  0.52 & CH           & $  1.5 \times 10^{ -9}$  &  0.83 \\
OCN          & $  1.7 \times 10^{-10}$  &  0.40 & HNC          & $  1.2 \times 10^{ -9}$  &  0.52 & Fe$^+$       & $  1.6 \times 10^{-10}$  &  0.84 \\
$^{\star}$OH & $  1.4 \times 10^{ -8}$  &  0.41 & H$_2$O$_2$   & $  4.0 \times 10^{ -9}$  &  0.53 & Mg$^+$       & $  3.3 \times 10^{-10}$  &  0.84 \\
P            & $  1.5 \times 10^{-11}$  &  0.41 & C$_4$H$_2$   & $  2.4 \times 10^{-11}$  &  0.53 & O$_2$H       & $  1.3 \times 10^{-11}$  &  0.85 \\
HC$_3$O      & $  3.2 \times 10^{-11}$  &  0.41 & C$_2$H$_4$   & $  1.7 \times 10^{-11}$  &  0.54 & CN           & $  1.4 \times 10^{ -9}$  &  0.86 \\
H$_2$S       & $  1.3 \times 10^{ -9}$  &  0.41 & C$_5$        & $  1.1 \times 10^{-10}$  &  0.54 & HC$_3$N      & $  1.6 \times 10^{-11}$  &  0.86 \\
FeH          & $  2.4 \times 10^{-10}$  &  0.42 & C$_3$H$_2$   & $  8.7 \times 10^{-11}$  &  0.54 & C$^+$        & $  1.5 \times 10^{ -6}$  &  0.88 \\
C$_4$        & $  1.6 \times 10^{-10}$  &  0.43 & C$_5$H$_2$   & $  2.1 \times 10^{-11}$  &  0.55 & C$_9$H$_2$   & $  2.3 \times 10^{-11}$  &  0.94 \\
NO           & $  4.1 \times 10^{ -9}$  &  0.43 & $^{\star}$CH$_4$
                                                               & $  7.7 \times 10^{ -8}$  &  0.55 & C$_5$H$_4$   & $  2.9 \times 10^{-11}$  &  0.97 \\
NH           & $  9.6 \times 10^{-11}$  &  0.43 & CH$_3$OH     & $  9.3 \times 10^{-11}$  &  0.56 & C$_2$        & $  1.2 \times 10^{ -9}$  &  1.02 \\
S            & $  6.5 \times 10^{ -9}$  &  0.43 & C$_5$H       & $  2.2 \times 10^{-11}$  &  0.57 & CH$_2$NH$_2$ & $  4.8 \times 10^{-11}$  &  1.09 \\
SiO          & $  3.4 \times 10^{-10}$  &  0.43 & CH$_2$OH     & $  4.1 \times 10^{-11}$  &  0.59 & CH$_3$NH     & $  4.7 \times 10^{-11}$  &  1.09 \\
$^{\star}$O$_2$ 
             & $  2.4 \times 10^{ -6}$  &  0.44 & HCO$^+$      & $  8.7 \times 10^{-11}$  &  0.60 & CH$^+$       & $  3.7 \times 10^{-13}$  &  1.19 \\
HNCO         & $  1.9 \times 10^{ -9}$  &  0.44 & HCOOH        & $  1.3 \times 10^{-11}$  &  0.63 & N$_2$H$_2$   & $  6.9 \times 10^{-11}$  &  1.22 \\
\enddata
\end{deluxetable*}

\begin{figure*}[ht!]
\plotone{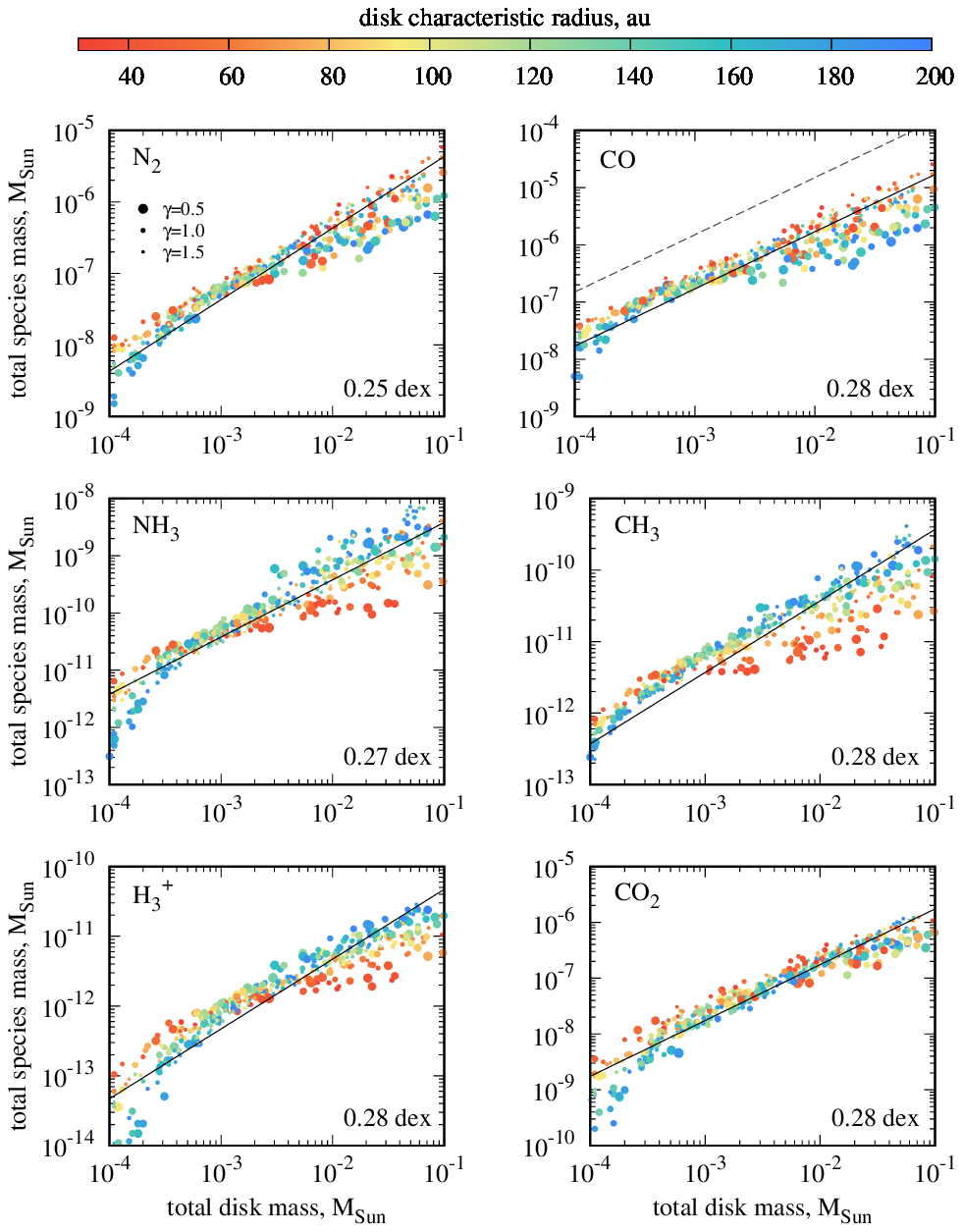}
 \caption{Total mass of species with lowest scatter values in the disk as a function of the disk mass at an age of 3~Myr. Point size depicts 
$\gamma$-value (larger symbols represent shallower surface density profiles), point color represents disk characteristic radius $R_{\rm c}$. Solid lines 
indicate linear scaling with the slope equal to average mass fraction. Dashed line corresponds to CO amount if it would contain all the available carbon 
atoms. The scatter parameter is shown in the right bottom corner of each panel.}
  \label{fig:confetti1}
\end{figure*}

The top of the table is occupied by N$_2$, NH$_3$, CH$_3$, CO, H$_3^+$, and CO$_2$, and in Fig.~\ref{fig:confetti1} their total masses are shown as 
functions of the total disk mass. In each panel a solid line is shown with the slope corresponding to linear scaling of the species mass with the 
disk mass at the average total abundance $\overline{X_i}=10^{\overline{x_i}}$ from Table~\ref{tab:variance}. A scatter parameter is also indicated in each panel. 
Molecular nitrogen seem to be the best {\em theoretical} mass tracer as its mass scales almost linearly with the disk mass over the entire range of 
considered disk masses. This molecule, like H$_2$, has no rotational transitions and thus cannot be observed directly, but $^{14}$N$^{15}$N has. Despite the mass difference between $^{14}$N and $^{15}$N is small and hence the dipole moment is very weak, theoretically, one could detect its fundamental
rotational line at 4.34~$\mu $m in disks.

The CO total mass also shows nearly linear scaling with the disk mass, but with a scatter that increases somewhat in more massive disks. Disks with lower $\gamma$ values (shallower density profiles) tend to have smaller CO abundance, while more compact disks (with 
lower $\gamma$ and $R_{\rm c}$, i.e. smaller and redder dots in Fig.~\ref{fig:confetti1}) are more CO rich. The reason is that in a compact disk most mass 
is located in a warm disk area, where CO freeze-out is prohibited. As we consider larger and/or shallower disks, more mass is shifted to colder disk areas.

The dashed line in the top left panel of Fig.~\ref{fig:confetti1} indicates the maximum possible CO mass, corresponding to the assumption that all carbon atoms are locked in gas-phase CO. In this extreme case $M_{\rm CO} = 1.5 \times 10^{-3} M_{\rm disk}$. According to our results, the computed total CO mass is much lower in all the considered models, and its typical value is only $1.7 \times 10^{-4} M_{\rm disk}$. Most carbon atoms are locked up not in CO but in CO$_2$ ice. Carbon redistribution is discussed in more detail in Section~\ref{sec:carbon}.

The scatter parameter for NH$_3$ and CH$_3$ is formally somewhat lower than the one for CO. However, inspection of Fig.~\ref{fig:confetti1} shows that 
masses of these species depend stronger on the disk structure, than the CO mass, if the disk mass exceeds $\sim10^{-3}\,M_{\odot}$. Also, ammonia masses 
loose linear scaling with disk masses in the least massive disks. Additionally, abundances of NH$_3$ and CH$_3$, as well as the one of H$_3^+$, are quite low, which hampers their detection. While NH$_3$ has nevertheless been detected by \textit{Herschel} in nearby TW~Hya disk \citep{2016A&A...591A.122S}, methyl radical and H$_3^+$ have no allowed dipole moment as they are symmetric and planar.
Potentially H$_3^+$ can be traced by o-H$_2$D$^+$, thus, even though we do not consider isotopic chemistry, o-H$_2$D$^+$ may also correlate well with the disk 
gas mass. However, detecting this species in disks is challenging even for ALMA \citep{2011A&A...533A.143C}.

Carbon dioxide looks quite promising. Its scatter only slightly exceeds that of CO and it will also be observable by JWST. While the dependence of 
its mass on the disk mass is non-linear in disks with the lowest masses, but this is compensated by uniformly low scatter and linear scaling for disks 
with masses above $\sim10^{-3}\,M_{\odot}$.

\begin{figure*}[ht!]
\plotone{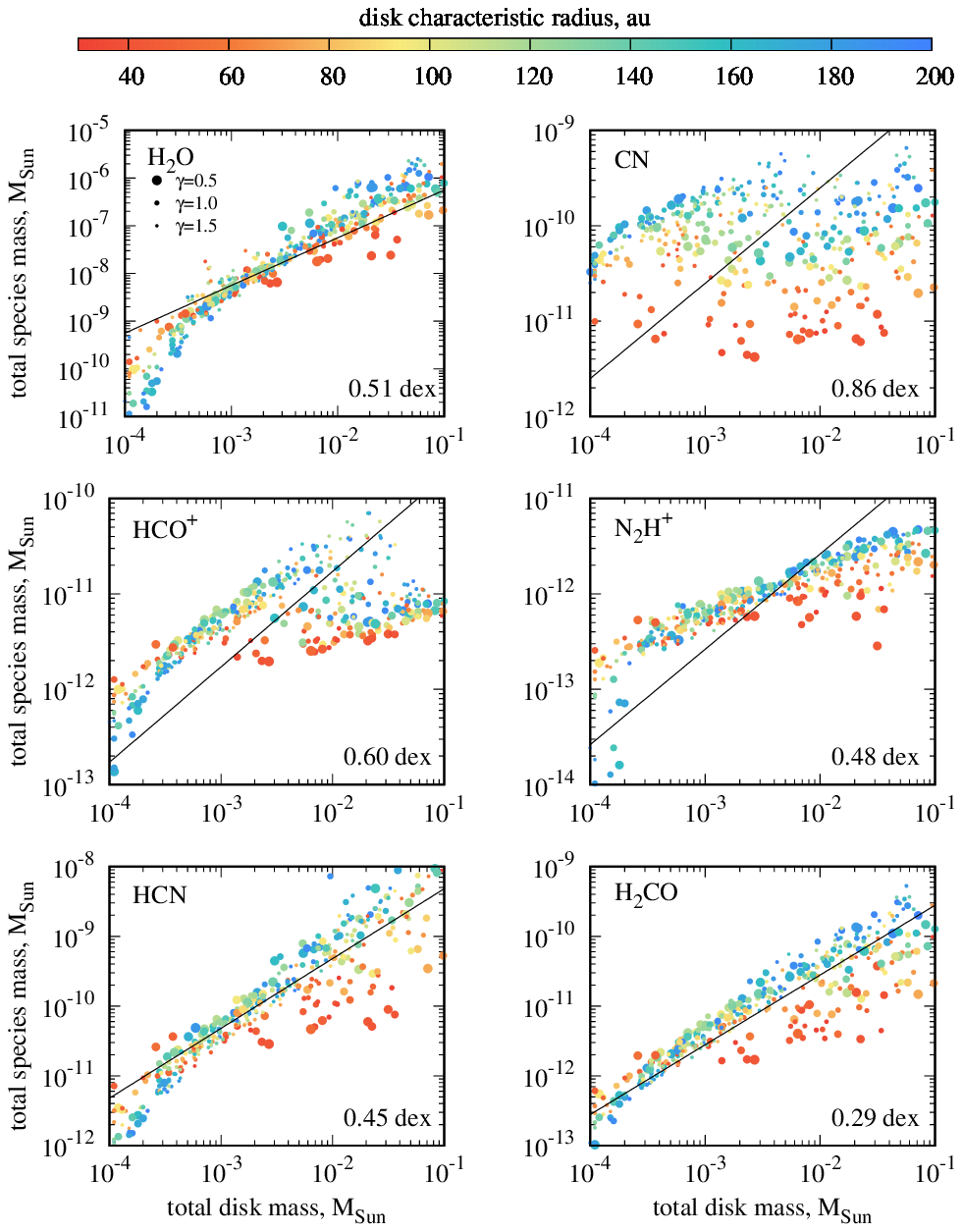}
 \caption{Same as in Fig.~\ref{fig:confetti1}, but for species with higher scatter values.}
  \label{fig:confetti2}
\end{figure*}
Overall, we conclude that currently CO is the best mass tracer combining ease of detection and predictability of behavior.
Among other species some are better than others, with the scatter comparable to the one of CO (see Fig.~\ref{fig:confetti2}). Water mass scales 
with the disk mass for $M_{\rm disk}>10^{-3}\,M_{\odot}$, but in the least massive disks water abundance shows a significant scatter (of about two orders 
of magnitude), with smaller $\gamma$ values and more extended disks corresponding to lower water masses. This seems to be interesting in a view of diverse 
results of {\em Herschel} water observations in disks of DM Tau \citep{bergin_dmtau} and TW Hya \citep {hoger_twhya}. Quite different water abundances may 
reflect not only evolutionary changes, but also structural variations.

Other carbon-bearing species show a significant scatter across the entire disk mass range, and the scatter generally increases as the disk mass 
increases. But in some cases the uncertainty can be reduced significantly if the disk size is known. Intriguingly, the H$_2$CO emission detected in disks could potentially be used as a proxy of disk mass, since the H$_2$CO-disk mass factor has a low scatter similarly to CO and is quite tight (albeit non-linear), if most compact disks are excluded. On the other hand the H$_2$CO emission in disks shows peculiar ring-like structure \citep{2008IAUS..251...89H, 2015ApJ...809L..25L, 2017ApJ...839...43O}, which makes such an analysis  a difficult endeavor. The CN and HCO$^+$ scaling is sufficiently non-linear that hampers their usage as mass tracers. 

An interesting case is represented by N$_2$H$^+$. Dependence of its mass on the disk mass is quite tight over the entire $M_{\rm disk}$ range and shows 
extreme sensitivity to the disk mass for $M_{\rm disk} < 3 \times 10^{-4}\,M_{\sun}$. However, at higher $M_{\rm disk}$ values the dependence of N$_2$H$^+$ mass on the disk mass is much weaker. This should be related to the CO behavior. In extremely low-mass disks CO is depleted from the gas phase only due to freeze out. In more massive disks CO also experiences the chemical depletion due to carbon redistribution into other species. The steeper left part of the N$_2$H$^+$ graph (see Fig.~\ref{fig:confetti2}) reflects the developing chemical depletion of CO, while at the shallower right part this process is saturated. The change of N$_2$H$^+$ abundance in more massive disks is explained by the 
shift of N$_2$H$^+$ from the midplane to upper layers.

\subsection{Carbon redistribution}
\label{sec:carbon}

\begin{figure}
\plotone{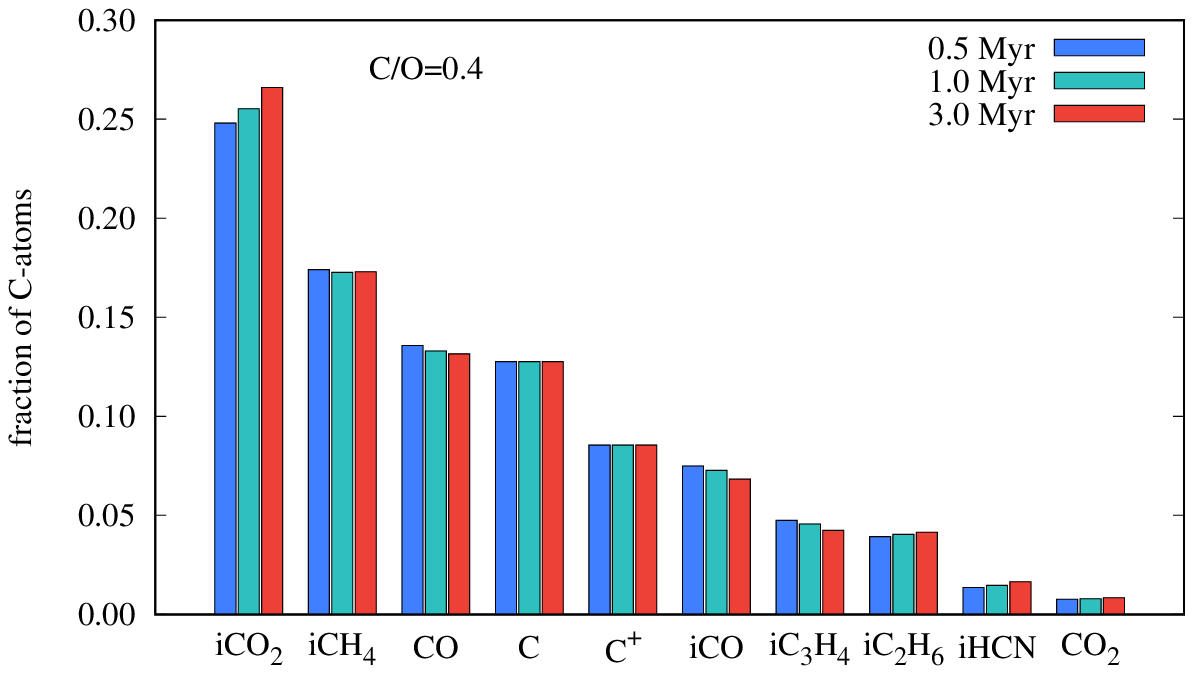}
 \caption{Distribution of carbon between major carbon-bearing molecules for different disk ages, averaged over the models. Ice species are denoted by 
leading `i'. C/O=0.4 is a ratio of carbon and oxygen elemental abundances.}
  \label{fig:histogram}
\end{figure}

In this section we consider major carbon-bearing species and demonstrate that gaseous CO is not the main reservoir of carbon in the considered 
protoplanetary disks. For each disk we find molecules which are major carbon reservoirs and calculate what percentage of carbon they comprise. The 
distribution of carbon among top-ten carbon-bearing species is plotted in Fig.~\ref{fig:histogram}. Three time moments are presented, 0.5,~1, and 3~Myr, 
showing that this distribution does not change much with the disk age.

There are two reasons for such a behavior. First, we show carbon fraction integrated over the entire disk, and most of material contributing 
to this value is relatively warm and dense, allowing chemical equilibrium to be reached within 0.5~Myr. Second, shown ice species are relatively simple, 
and their formation does not involve long-scale surface chemistry, so that they also have short typical chemical timescales.

We do see, however, some temporal changes in the carbon fraction for some species. The fraction of CO$_2$~ice gradually 
increases with time, while fractions of CO in gas and ice diminish mostly due to conversion of iCO into iCO$_2$ on the dust surfaces. Other molecules 
fractions also demonstrate some trends, so the redistribution of carbon in disks slowly continues during the considered time.

The diagram in Fig.~\ref{fig:histogram} shows that carbon is mostly locked in CO$_2$-ice (27$\pm$8\%) at 3~Myr, but a sufficient fraction of 
carbon stays in elementary forms of C or C$^+$. Among ten major carbon-bearing molecules, which constitute 95\% of all available C~atoms, only 
four are in the gas phase. The most abundant gas-phase species is CO with an average of 13\%, while gas-phase CO$_2$ comprises only about 
1\%. In other words, only every seventh C atom belongs to CO in a gas form. The prevalence of the CO$_2$ ice, which is mostly formed on 
grains, stresses the necessity of proper surface chemistry treatment.

\begin{figure*}
\plotone{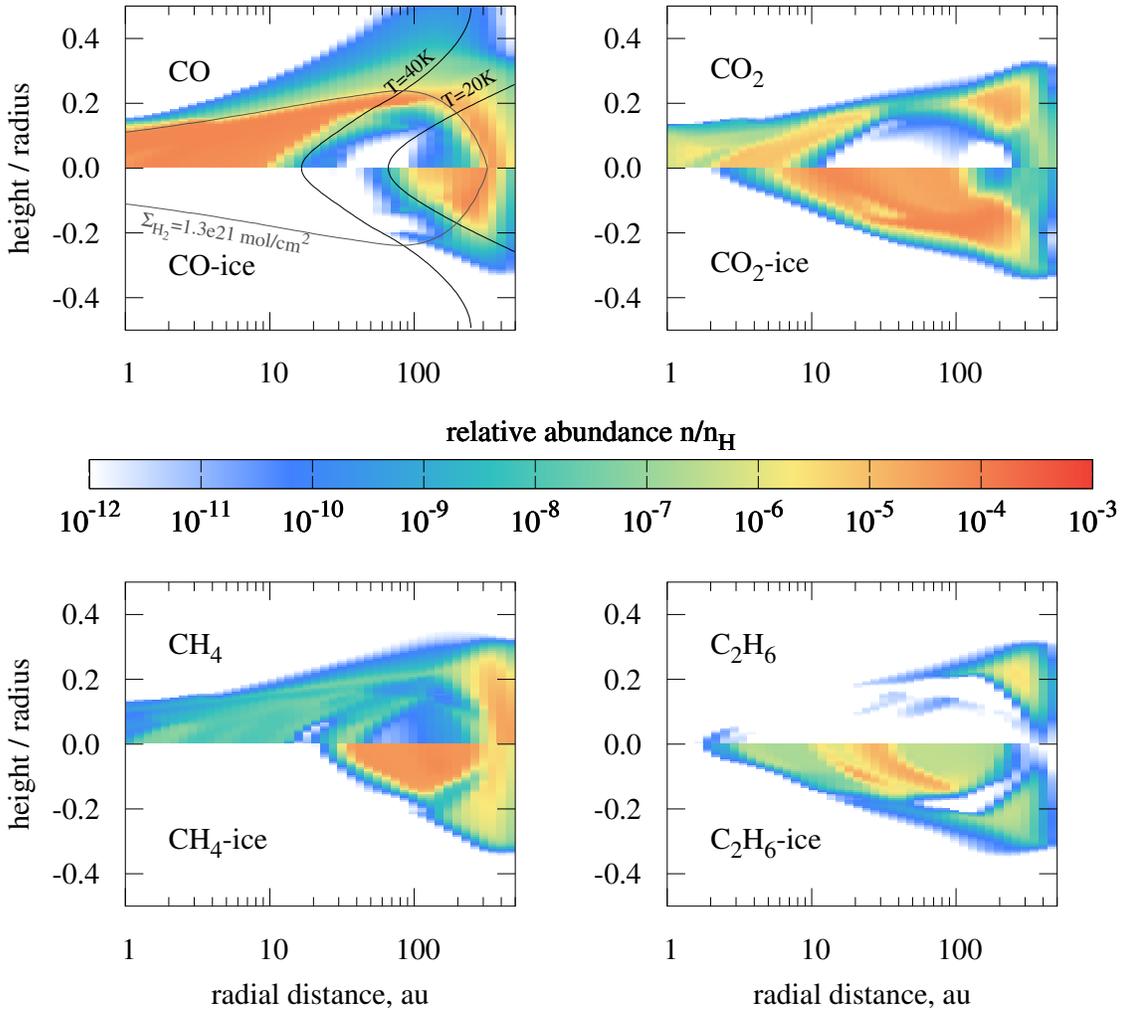}
 \caption{Abundances of major carbon-bearing species over the disk at the age of 3~Myr, for one particular model ($M_{\rm disk} = 0.01~M_{\sun}$, 
$M_{\star} = 1.0~M_{\sun}$, $R_{\rm c} = 100$~au, and $\gamma = 0.8$). Upper half of the plots corresponds to gas abundances, while the bottom half shows 
ice abundances.}
  \label{fig:carbomolecules}
\end{figure*}

In Fig.~\ref{fig:carbomolecules} we present distribution over the disk of important C-containing molecules for a model with typical parameters ($M_{\rm 
disk} = 0.01~M_{\sun}$, $M_{\star} = 1.0~M_{\sun}$, $R_{\rm c} = 100$~au, and $\gamma = 0.8$). Despite CO snowline is expected to be located at 
$\sim 70$~au, where the temperature reaches the critical value of 20~K, we see lack of gaseous CO in the disk midplane as close to the star as 
$\sim$15~au. Even within the CO snowline in a region with temperature just above 20~K there is constant exchange of CO molecules between gas and 
ice reservoirs. Here some of temporally frozen-out CO molecules are apparently converted into CO$_2$ ice before they go back to the gas. Black and gray 
contours in Fig.~\ref{fig:carbomolecules} indicate the area, limited by the criteria outlined by \citet{WB2014}: CO is supposed to reside in the gas-phase 
everywhere except for the regions where it is frozen out ($T < 20$~K) or photo-dissociated ($\Sigma_{\rm H_2} < 1.3 \times 10^{21}$~cm$^{-2}$). But in our 
model CO is not only highly depleted inside the restricted area due to chemical redistribution into other molecules, but it is also present outside of 
this area. Specifically, we see some CO in the outer disk region, which is shadowed from the star. In this cold region CO is photo-desorbed from dust grains, 
but not photo-dissociated by interstellar UV due to H$_2$ shielding. This allows CO to be present in the gas phase despite temperatures below the freezing 
point.

In our set of models average total abundance ratio iCO$_2$/iH$_2$O is 1/3. For the model shown in Fig.~\ref{fig:carbomolecules} the 
iCO$_2$/iH$_2$O abundance ratio varies with radius from 10\% to 110\% in the 6--120~au range with maximum at 20~au. CO-ice dominates over CO$_2$-ice 
beyond 200~au with maximum abundance ratio relative to water of 60\% at 300~au. These characteristic abundance ratios are consistent with what is 
observed in most comets \citep{mumma}. However, comparing our results to cometary abundances, one has to keep in mind that abundances of molecules 
observed 
in comets depend also on the nuclei structures and their thermal evolution \citep[e.g.][]{hassig2015time} and on the molecule processing in comae. The 
dynamical evolution of the comet ensemble is also of importance. In our model this ratio is computed for a large range of models and may differ from the 
corresponding ratios in specific models and/or spatial locations.

The CO$_2$ snowline is clearly seen at $\sim$6~au, and a significant amount of CO$_2$ ice resides in the disk dark regions up to distances of 
about 300 au from the star. Another abundant ice, methane, has a snowline at about 20--30~au from the star and is quite abundant in the midplane 
from 30 to 300~au. The region of abundant ethane ice extends down to a few au to the star, while in the gas-phase this molecule is nearly absent. 
Overall, among all the carbon-bearing species CO is not the most abundant, but it is the only species, which resides mostly not in ices, but in the 
gas-phase.

Lines defined by the above criteria delineate the part of the disk that contains 56\% of its mass for the presented model. In other disks from our 
sample this value changes from less than 5\% in extremely low-mass disks transparent for UV photons and up to 94\% in warm compact 
disks, with the mean value of about 50\% over the whole ensemble. If the ISM-like CO abundance of $10^{-4}$ relative to H$_2$ were 
present everywhere in the disk, and assuming that CO traces on average only 50\% of the disk mass, we would get CO average abundance in disk equal to $ 5 
\times 10^{-5}$. On the other hand, our chemical modeling suggests $\overline{X_{\rm CO}} = 8 \times 10^{-6}$ (see Table~\ref{tab:variance}). 
Thus, we conclude that CO/H$_2$ abundance ratio of $1.6 \times 10^{-5}$ should be used for determining the disk mass from CO observations, and qualitative 
criteria of CO depletion overestimate its presence in the disk by a factor of~3.

Another way to improve \cite{WB2014} formalism would be to apply a different value of critical temperature below which CO is absent in the gas phase. It 
can be seen from Fig.~\ref{fig:carbomolecules} that the abundance of CO does reach $10^{-4}$ in part of the disk where it is predominantly in the gas 
phase. The photo-dissociation limit is traced as well, it only misses gaseous CO in the region of low density in the disk atmosphere far from the star, 
which makes an order of a few percent contribution to the total CO mass. However, the line of CO freeze-out defining the CO-gas region should be replaced 
by the chemical depletion front, coincident with the $T\approx40$~K isoline (see Fig.~\ref{fig:carbomolecules}).

\subsection{Role of individual parameters}
\label{sec:gridofmodels}

\begin{figure}
\plotone{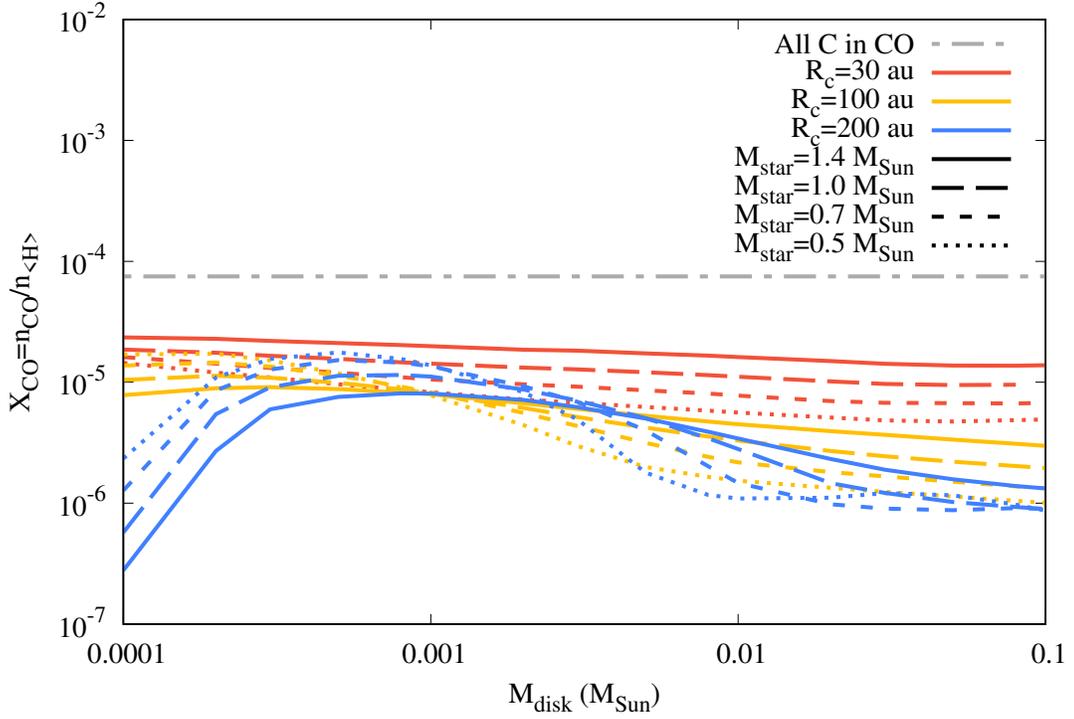}
 \caption{Total disk abundance of CO in the disks of different characteristic radii $R_{\rm c}$ and around stars of different masses, $\gamma = 
0.5$, at the age of 3~Myr. Line color indicates $R_{\rm c}$, line styles (solid, dashed and dotted) indicate star mass. Maximum $X_{\rm CO}$ is shown with 
the gray line.}
  \label{fig:grid}
\end{figure}

In this section, we use the grid of models to investigate how the various parameters influence the CO abundance in the disk individually. For that purpose, 
we fix some parameters and check how the results change due to variations in other parameters.

Fig.~\ref{fig:grid} illustrates the influence of the disk radius and the stellar mass onto the scatter of CO total abundance at a fixed value 
$\gamma = 0.5$. The star mass has only a small impact on the derived CO abundance in compact disks ($R_{\rm c}=30$ au). When we vary $M_{\star}$ 
between 0.5~and 1.4~$M_{\sun}$, the CO total abundance in compact disks varies only by about a factor of 3, and the higher the star mass is, the 
larger the CO abundance is. The influence of the star mass becomes more dramatic in disks of larger radii, at least, at the lower end of the considered 
disk mass range. Here the same variation in $M_{\star}$ leads to an order of magnitude difference in $X_{\rm CO}$. The overall 
uncertainty of the CO content reaches two orders of magnitude in the least massive disks.

It is also interesting to note that in models with lower disk masses a hotter star leads to a smaller CO abundance, while in models with higher disk masses 
the opposite trend is observed. Obviously, the CO abundance is limited by the photo-dissociation and photo-desorption, and in various situations the higher 
star temperature can lead to either decrease or increase in the gas-phase CO abundance. Very low-mass disks, being more transparent for stellar 
UV-radiation, suffer from CO photo-destruction, which is stronger in disks around more massive stars. Thus, these disks contain less gas-phase~CO. Low-mass 
disks with $R_{\rm c}=30$~au do not show this trend because they encompass the same mass in a more compact region, which results in higher densities and 
lower influence of photo-dissociation. In massive disks photo-dissociation is not that crucial, and CO depletion is mostly caused by chemical 
redistribution into ices, especially CO$_2$~ice. A less massive star causes a CO$_2$ snowline closer to the star, expanding the zone where CO is 
effectively removed by chemical processes. 

Other considered values of $\gamma$ (0.8, 1.0 and 1.5) do not produce such a high variance in CO total abundance, with larger $\gamma$ giving more stable result. The differences described above, are the highest in disks with $\gamma = 0.5$, where surface density falls off slowly, leaving more material in the outer regions of the disk, where CO depletion is stronger. The dependence on $R_{\rm c}$ is caused by a similar effect: disks of equal mass 
have more matter in the warm inner region if their radii are small, allowing more CO to reside in the gas phase. When $R_{\rm c}$ is large, more mass is in 
the cold outer area with high CO depletion.

\subsection{Initial C/O ratio}

\label{sec:agenc2o}

\begin{figure}
\plotone{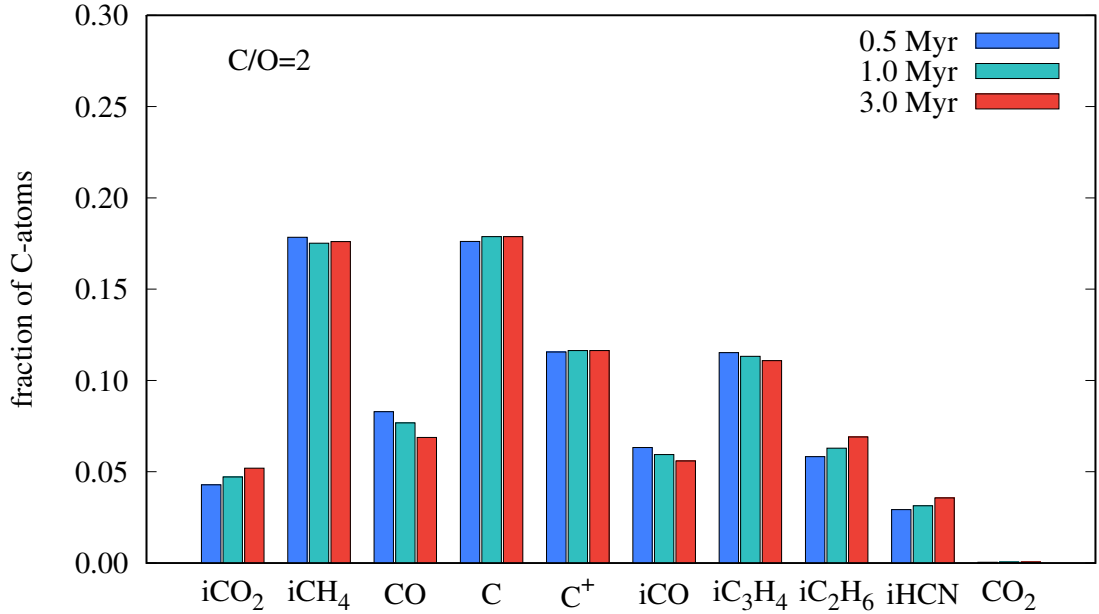}
 \caption{Same as Fig.~\ref{fig:histogram}, but for C/O~=~2.}
  \label{fig:histogram2}
\end{figure}

The spatial distribution of carbon described in Section~\ref{sec:carbon} was computed for a disk with low-metallicity initial composition 
(Table~\ref{tab:iniab}), where there is more oxygen than carbon (C/O$\approx$0.4). To test the effect of other possible elemental compositions we 
run $200$ models, reducing the initial O abundance from $1.76\times10^{-4}$ to $3.65 \times 10^{-5}$ (leading to C/O~=~2).

Fig.~\ref{fig:histogram2} presents the fraction of carbon atoms locked up in the various species, dominant in the disk at the previous value of C/O. 
Compared to Fig.~\ref{fig:histogram}, the abundances of CO$_2$ and CO both in the gas and ice dropped significantly because of the lack of oxygen. On average, 
only $\sim$7\% of C is now in CO gas, while the amount of CO~ice stays nearly the same.

Instead of CO$_2$ ice we have a richer carbon chemistry on dust surfaces. In addition to ices iCH$_4$, iC$_2$H$_6$ and iC$_3$H$_4$ we end up with much 
higher abundances of molecules like iC$_5$H$_2$, iCH$_3$CN, iC$_9$H$_2$, iCH$_3$C$_3$N and many other surface species with long C~chains 
\citep{2009ARA&A..47..427H}. Abundances of atomic and ionized carbon are also higher compared to the case of C/O=0.4.

Overall, the carbon distribution among species and the abundance of CO depend on initial elemental composition as well.

For the present set of models we have calculated the scatter parameter $s$ as well. The CO scatter parameter is still close to the top with the 
value of 0.22, it goes again after N$_2$ with $s=0.20$, and few minor species having the same values of scatter but low mean abundances ($<10^{-9}$). For 
this elemental composition CO$_2$, previously showing a low scatter, has $s=0.60$, which worsens its applicability as a mass tracer. Among top species 
only CO and N$_2$ retain approximately the same values of scatter at C/O=2.

\section{Discussion}
 
While observations of molecular lines remain the only tool to determine gas masses of protoplanetary disks, their interpretation is far from being easy. 
Even though the most common mass tracer in the interstellar and circumstellar medium, CO, is believed to be controlled by a relatively simple set of 
processes, its straightforward application to gas mass determination in protoplanetary disks has produced results, which contradict both dust 
observations and HD observations \citep{2013Natur.493..644B,2013ApJ...776L..38F,2016ApJ...831..167M}. The discrepancy between dust-derived and CO-derived 
disk masses can in principle be explained by the dust evolution and gas dispersal, but it is harder to reconcile data on various molecular gas mass 
tracers, like CO and HD, though this difference can be reduced by accounting for CO isotope selection and carbon underabundance in disks, as 
\citet{2017arXiv170507671T} suggest. 

The basic assumption behind the possibility of using CO as a mass tracer is that its abundance in protoplanetary disks is defined by the balance between 
photo-dissociation and freeze-out. \cite{WB2014} suggested a grid of models based on the assumption that CO is frozen out everywhere, where the 
temperature is below 20~K, and is photo-dissociated above the H$_2$ column density of $1.3\times10^{21}$~cm$^{-2}$. If none of these conditions is met, CO 
is assumed to have the `interstellar' abundance of $10^{-4}$. It becomes increasingly clear that simple prescriptions, like the one suggested by 
\cite{WB2014}, can produce spurious results. First, as \cite{WB2014} already noted, more efficient depletion pathways are possible rather than simple CO 
ice formation. Also, the CO sublimation temperature depends on pressure and under conditions of the disk midplane could be higher than 20~K 
\citep{2015A&A...582A..41H}\footnote{There is also evidence for the opposite. In some cases, CO in the midplane is detected at temperatures as low as 13~K 
\citep{2003A&A...399..773D}, which, probably, can be explained by turbulent mixing \citep{2006ApJ...647L..57S}.}. Second, a typical `interstellar' CO 
abundance (inherited by the disks) can be lower than $10^{-4}$ \citep{2007ApJ...658..446B}. One way to infer CO/H$_2$ ratio in protoplanetary disks 
directly is to observe absorption lines of both components in the UV band \cite{france}. This can be one of prospectives for future space UV missions like 
WSO-Spectrum UV \citep{2016ARep...60....1B} and LUVOIR\footnote{\url{https://asd.gsfc.nasa.gov/luvoir/}}.

Obviously, more sophisticated methods should be used to infer the total gas mass from CO (or other molecule) observations. In a number of recent papers the 
utility of CO as a mass tracer has been assessed by means of detailed models of the disk physical and chemical structure. In a series of works by 
\cite{2014A&A...572A..96M,2016A&A...594A..85M,Miotello2017} a disk chemical model with isotope-selective processes was considered in order to find a way to 
determine the disk mass from CO isotopologue line observations. A large set of disk models was employed to study the evolution of $^{13}$CO, C$^{17}$O, and 
C$^{18}$O in these disks. It was concluded that CO observations can lead to underestimating gas masses in those disks where dust grains have grown to 
larger sizes. Application of the results to real observations produced very low gas masses in T~Tauri disks, often lower than 1 $M_{\rm J}$. It was noted 
that this can be an effect of sequestering carbon atoms to more complex molecules. This option has not been considered in the above study, as the 
used chemical model included only a limited set of surface processes (simple hydrogenation processes). The consequence can be seen in Figure~2 from 
\cite{2014A&A...572A..96M}, where the CO depletion zone is delineated by a $T=20$~K line, and the CO snowline is located at about 100~au.

A more detailed consideration of surface processes was performed by \cite{2015A&A...579A..82R}. A single surface density distribution was considered in 
this study, but with different vertical temperature profiles. It was shown that surface conversion of CO into other ices, primarily CO$_2$ ice, leads to 
gas-phase CO abundances much lower than the canonical value of $10^{-4}$. Abundances that high are only reached in relatively warm disks with midplane 
temperatures above 30~K beyond $R=100$~au. In this work results were presented only for the outer part of the disk ($R=100$ and 300~au), where theoretical 
CO abundance can be quite sensitive to details of radiation transfer \citep[see our results above and also observations presented 
in][]{2016ApJ...823L..18H}.

A chemical evolution with a detailed account of surface reactions and carbon isotope-selective processes was considered by \cite{Yu2016,Yu2017}. They 
studied the chemical evolution of a typical disk within inner 70~au and found that due to chemical depletion, i.e. conversion of CO into less volatile 
molecules, the effective CO snowline is located at about 20~au and moves toward the star as the disk evolves, even though the temperature in their model is 
above the CO sublimation point everywhere in the considered region. According to their results, after 3 Myr of evolution 13.6\% of all available carbon is 
locked in gas-phase CO, while CO$_2$ ice contains 36.2\% of carbon atoms, which compares favorably with our results for a similar disk model (we 
must note that this agreement is reached despite quite different assumptions on dust properties). While the average value of carbon partition matches well, 
its dispersion is significant in our modeling (from a few per cent up to 50\%). They also found that CO abundance varies with time significantly, which 
is something that we do not see in our model. However, in their model they vary stellar properties with time, while we keep them constant. The overall 
conclusion of \cite{Yu2017} is that straightforward interpretation of CO observations leads to underestimation of disk masses. Thus, all the conclusions on 
the gas deficit in protoplanetary disks based on CO observations should be considered with caution.

The focus of the above studies was on isotope-selective processes and on the usage of CO isotopologue line ratios as mass indicators. In our study we 
consider CO as a whole, without distinguishing between its isotopologues, concentrating rather on global chemical aspects of CO distributions in disks 
having different structural parameters.

In our study we consider a wide range of disk sizes, masses, and surface density profiles, assuming that there is no dust evolution, so that dust still 
has its interstellar parameters. Our results indicate that CO is not an ideal mass tracer, but all other molecules show greater uncertainties in relative 
abundance. Still, some of them, e.g. H$_2$O, H$_2$CO and especially CO$_2$ can be used as supplementary tracers, particularly, if some 
information on the disk structure is available.

A better calibration of CO observations can be obtained using the \cite{WB2014} method, assuming a factor of a few lower typical abundance and taking into 
account the chemical depletion (e.g., taking 40~K as CO depletion temperature). 

An obvious extension of the study is to consider the dependence of our results on the assumed grain size as dust is expected to grow in protoplanetary 
disks. \cite{2013ApJ...766....8A} have shown that the effect of grain growth at its initial stage is to shift the molecular layer closer to the dense 
midplane. We may provisionally expect that this is going to make molecules better gas mass tracers, but this is a subject for the future study.

\section{Conclusions}

We conducted a chemical modeling of $\sim 1000$ protoplanetary disk structures around low-mass stars with different parameters to find out species that 
have the best correlation with the total disk mass. We varied the central star mass, the disk characteristic radius, the disk mass, and 
$\gamma$-index of surface density distribution, assuming MRN grain size distribution for the UV radiation transfer in the disk atmosphere and 0.1$\mu $m-size grains for the surface chemistry. The chemical modeling is based on the updated non-equilibrium 
chemical network ALCHEMIC \citep{2011ApJS..196...25S} incorporated into the ANDES code \citep{2013ApJ...766....8A}. We focused on an intrinsic dispersion 
of species abundance due to unknown structural and thermal parameters. The main conclusions are:

-- among all 650 considered species, the relative abundance of the CO molecule has one of the smallest scatter in the overall disk ensemble (with 
obvious exceptions of hydrogen and helium). The characteristic reference `$1\sigma$'-values for the logarithm of species total disk abundance at an age of 
3~Myr are 0.28~dex for CO and CO$_2$, 0.29~dex for H$_2$CO, 0.51~dex for H$_2$O (see Table~\ref{tab:variance} for other species). So, 
even in the case of CO, there is a maximum uncertainty of 1 order of magnitude in CO abundance if the disk physical structure is unknown. 
More reliable CO disk masses can be obtained if the disk characteristic radius $R_{\rm c}$ is determined.

-- on average in the whole disk ensemble, $\approx$ 13\% of carbon atoms end up in the gas-phase CO molecule (for the initial `low-metals' 
abundance from \cite{1998A&A...334.1047L} where C/O~=~0.4). The average value of carbon partitioning is in a good agreement with conclusions by 
\citet{Yu2016}. However, the extreme values vary from $\approx$ 2\% for large and massive disks to $\approx$ 30\% for compact disks with small 
and moderate masses.  Statistically, the most abundant carbon-bearing species is CO$_2$-ice with average C-atom fraction of $\approx 25\%$. The 
majority of the remaining carbon atoms are in CH$_4$ and CO ices (17\% and 7\% of C-atoms, respectively), C and C$^+$ (14\% and 8\%) and 
ices of complex organic species. The degree of carbon atom redistribution into non-CO molecules is naturally more effective if C/O~$>1$. Quantitatively, 
the average carbon partition into CO is 7\% for C/O~=~2.

-- despite CO has a noticeable variance in the relative abundance, it is still the best molecular tracer of disk gas mass. The typical 
value of total CO/H$_2$ abundance ratio is $1.6\times10^{-5}$ with `$1\sigma$'-limits from $0.8\times10^{-5}$ to $3.0\times10^{-5}$. CO$_2$ also has a low abundance variance, though this variance depends strongly on C/O ratio. Other species that have relatively good correlation with disk 
mass are H$_2$O and H$_2$CO.

We find that on average the total abundance of gaseous CO in protoplanetary disks is $\approx$~6 times lower compared to the interstellar value of $10^{-4}$. Chemical depletion lowers the abundance of CO by a factor of 3, compared to the case of photo-dissociation and freeze-out as the only ways of CO 
destruction. 

\acknowledgements
We thank the referee for fruitful comments that helped us to improve our model and to formulate better the conclusions.
TM, VA and DW acknowledge financial support from Russian Science Foundation (17-12-01441; Sections~2, 3, 5). DS acknowledges support from the Heidelberg 
Institute of Theoretical Studies for the project ``Chemical kinetics models and visualization tools: Bridging biology and astronomy''. AV acknowledges 
support from the European Research Council (ERC; project PALs 320620).

\appendix

\section{Ice species}
Here we present the version of Table~\ref{tab:variance} for ice species. Table~\ref{tab:icevariance} lists ice species with average total 
abundance in the disk above $10^{-11}$ sorted by the scatter.

\begin{deluxetable*}{l|ll|l|ll|l|ll}
\tablecaption{Average total abundance $\overline{X_i}=10^{\overline{x_i}}$ and scatter $s_i$ for ice species\label{tab:icevariance}.}
\tablewidth{500pt}
\tablehead{
\colhead{Species} & \colhead{$\overline{X_i}$} & \colhead{$s_{i}$, dex} & 
\colhead{Species} & \colhead{$\overline{X_i}$} & \colhead{$s_{i}$, dex} & 
\colhead{Species} & \colhead{$\overline{X_i}$} & \colhead{$s_{i}$, dex} }
\startdata
iSiO           & $  8.1 \times 10^{-10}$  &  0.20 & iCH$_3$CN    & $  1.1 \times 10^{ -9}$  &  0.62 & iC$_6$H$_6$    & $  2.1 \times 10^{-11}$  &  0.94 \\
iCO$_2$        & $  1.7 \times 10^{ -5}$  &  0.26 & iC$_3$H$_4$  & $  6.7 \times 10^{ -7}$  &  0.65 & iH$_2$S$_2$    & $  5.1 \times 10^{-10}$  &  0.96 \\
iFeH           & $  1.9 \times 10^{ -9}$  &  0.26 & iPO          & $  1.2 \times 10^{-11}$  &  0.65 & iH$_2$CO       & $  6.6 \times 10^{ -8}$  &  0.96 \\
iMgH$_2$       & $  4.4 \times 10^{ -9}$  &  0.28 & iHC$_3$N     & $  2.1 \times 10^{-11}$  &  0.65 & iH$_2$         & $  1.1 \times 10^{ -7}$  &  0.96 \\
iC$_2$H$_4$    & $  3.0 \times 10^{-10}$  &  0.28 & iC$_7$H$_2$  & $  1.9 \times 10^{-11}$  &  0.66 & iH$_2$C$_3$O   & $  4.6 \times 10^{-10}$  &  0.99 \\
iHNC           & $  1.1 \times 10^{ -8}$  &  0.29 & iC$_8$H$_4$  & $  1.0 \times 10^{ -9}$  &  0.66 & iCH$_3$C$_4$H  & $  8.0 \times 10^{-11}$  &  1.05 \\
iHNCO          & $  3.5 \times 10^{ -9}$  &  0.34 & iC$_4$N      & $  1.5 \times 10^{-11}$  &  0.67 & iNH$_2$CHO     & $  2.3 \times 10^{ -9}$  &  1.10 \\
iSiH$_4$       & $  3.8 \times 10^{ -9}$  &  0.35 & iH$_3$C$_5$N & $  3.2 \times 10^{-11}$  &  0.69 & iCH$_2$CO      & $  1.7 \times 10^{ -9}$  &  1.17 \\
iP             & $  7.6 \times 10^{-11}$  &  0.36 & iCH$_3$OH    & $  5.2 \times 10^{ -8}$  &  0.69 & iHNO           & $  3.3 \times 10^{ -9}$  &  1.29 \\
iHCN           & $  9.8 \times 10^{ -7}$  &  0.37 & iC$_8$H$_2$  & $  1.6 \times 10^{-11}$  &  0.70 & iO$_2$         & $  7.6 \times 10^{-10}$  &  1.32 \\
iHCl           & $  2.4 \times 10^{-10}$  &  0.38 & iH$_5$C$_3$N & $  2.2 \times 10^{-10}$  &  0.71 & iS$_2$         & $  4.0 \times 10^{-10}$  &  1.38 \\
iC$_5$H$_2$    & $  3.3 \times 10^{-10}$  &  0.39 & iC$_7$H$_4$  & $  1.9 \times 10^{ -9}$  &  0.72 & iNH$_2$        & $  2.3 \times 10^{-11}$  &  1.39 \\
iCl            & $  2.9 \times 10^{-10}$  &  0.40 & iCH$_4$      & $  5.3 \times 10^{ -6}$  &  0.73 & iNO            & $  1.1 \times 10^{-11}$  &  1.49 \\
iH$_2$O        & $  5.1 \times 10^{ -5}$  &  0.40 & iC$_5$H$_4$  & $  7.6 \times 10^{ -8}$  &  0.73 & iC$_4$H$_4$    & $  4.3 \times 10^{-10}$  &  1.50 \\
iNaH           & $  9.5 \times 10^{-10}$  &  0.43 & iC$_2$H$_6$  & $  7.3 \times 10^{ -7}$  &  0.74 & iCH$_3$OCH$_3$ & $  1.9 \times 10^{-11}$  &  1.52 \\
iNH$_3$        & $  6.6 \times 10^{ -6}$  &  0.46 & iN$_2$H$_2$  & $  2.3 \times 10^{-11}$  &  0.76 & iC$_2$H$_5$OH  & $  2.0 \times 10^{-11}$  &  1.53 \\
iH$_2$O$_2$    & $  5.2 \times 10^{ -8}$  &  0.49 & iH$_2$CS     & $  4.5 \times 10^{ -9}$  &  0.81 & iOH            & $  5.0 \times 10^{-10}$  &  1.56 \\
iC$_3$S        & $  7.3 \times 10^{-11}$  &  0.49 & iHCOOH       & $  1.6 \times 10^{ -8}$  &  0.82 & iCH$_3$CHO     & $  2.3 \times 10^{ -9}$  &  1.57 \\
iC$_6$H$_2$    & $  9.4 \times 10^{-11}$  &  0.49 & iC$_9$H$_4$  & $  2.2 \times 10^{ -9}$  &  0.83 & iCH$_3$        & $  3.1 \times 10^{-11}$  &  1.60 \\
iNH$_2$OH      & $  7.8 \times 10^{-10}$  &  0.53 & iCH$_5$N     & $  9.6 \times 10^{ -9}$  &  0.83 & iCH$_2$        & $  3.0 \times 10^{-11}$  &  1.64 \\
iN$_2$         & $  6.8 \times 10^{ -7}$  &  0.55 & iC$_9$H$_2$  & $  6.3 \times 10^{-11}$  &  0.84 & iCH            & $  2.2 \times 10^{-11}$  &  1.75 \\
iCO            & $  3.2 \times 10^{ -6}$  &  0.58 & iO$_3$       & $  2.7 \times 10^{ -9}$  &  0.85 & iO             & $  1.3 \times 10^{-10}$  &  1.75 \\
iH$_2$S        & $  2.3 \times 10^{ -8}$  &  0.59 & iHS$_2$      & $  2.3 \times 10^{-10}$  &  0.86 & iC             & $  2.2 \times 10^{-11}$  &  1.86 \\
iC$_3$H$_2$    & $  1.2 \times 10^{ -9}$  &  0.60 & iC$_6$H$_4$  & $  1.8 \times 10^{ -8}$  &  0.88 & iH             & $  2.4 \times 10^{ -9}$  &  1.98 \\
iC$_2$H$_2$    & $  9.5 \times 10^{-11}$  &  0.61 & iN$_2$O      & $  1.1 \times 10^{-11}$  &  0.92 &                &                          &       \\
\enddata
\end{deluxetable*}

\bibliographystyle{aasjournal}
\bibliography{refs}
 
\end{document}